\documentclass[aps,prl,twocolumn,superscriptaddress,showpacs]{revtex4}
\usepackage{graphicx}
\begin{document}

\title{Quantum size effects in adatom island decay}
\author{Karina Morgenstern$^1$, Erik L\ae gsgaard$^2$,
Flemming Besenbacher$^2$}

\affiliation{
$^1$ Institut f\"{u}r Festk\"orperphysik, Universit\"at Hannover, Appelstr.\ 2,
D-30167 Hannover, Germany, \\
$^2$ Department of Physics and Astronomy, and Interdisciplinary
Nanoscience Center (iNANO), University of Aarhus, DK - 8000 Aarhus
C, Denmark}

\begin{abstract}
The decay of hexagonal Ag adatom islands on top of larger Ag adatom
islands on a Ag(111) surface is followed by a fast scanning tunneling microscope. 
Islands do not always show the expected increase in decay rate with decreasing island size. 
Rather, distinct quantum size effects are observed where the decay rate decreases significantly 
for islands with diameters of 6 nm, 9.3 nm, 12.6 nm, and 15.6 nm. 
We show that electron confinement of the surface state electrons is responsible for this 
enhancement of the detachment barrier for adatoms from the island edge.
\end{abstract}


\maketitle


Thin metal films and small quantum dots have a wide range of
applications in modern technologies. Quantum size effects are found
to occur when the size of the  nano\-structure  becomes comparable
to the de Broglie wavelength of electrons confined within it. Such
confinement effects have profound implications on various nanoscale
physical properties.

The first observation of quantum size effects in individual islands
was made for Pb islands on Cu(111) \cite{hinch89}, where it was
shown that preferred heights occur for islands without quantum well
states close to the Fermi level \cite{otero02}. Also Ag and Pb thin
films on semiconductor substrates showed magic heights
\cite{smith96,altfelder97,huang98,gavioli99}. The detection of
quantum states was extended towards thinner films and a correlation
between the electronic property and the thickness of an individual
island was demonstrated \cite{su01}. It has been proposed that the
competition between quantum confinement, charge spilling and
interface-induced Friedel oscillation defines the existence of the
characteristic magic thicknesses \cite{zhang98}. Thus, in a so
called ''electronic growth'' mode for metals on semiconductors the
energy contribution of quantized electrons confined in the metal
overlayer determines the morphology of the growing film, and this
energy contribution prevails over the strain energy. This insight
was used to tune the surface reactivity of magnesium films towards
oxidation \cite{aballe04}. It was also observed that annealing
of the Pb overlayer on Cu(111) \cite{dil04} and the Ag overlayer on Fe(100)
\cite{man04} leads to magic layer thicknesses associated with quantum
size effects. All these observations originate from the confinement
of electrons normal to the surface and it was implicitly 
assumed that the extension of the macrostructure was too large in the
lateral direction to produce lateral confinement. 
However, electron confinement of the
surface state electrons in the lateral dimension leads to quantum
interference patterns on nanoscale islands.
These patterns were observed
in both homoepitaxial \cite{li98} and heteroepitaxial \cite{pons01} systems. 

During the initial growth of adatom islands the system is out of
equilibrium since islands nucleate and since it cost free energy to form
the step edges. When nucleation and growth are terminated, atoms flow
from steps with high curvature to steps with low curvature making
small islands shrink, while large islands grow at their expense.
This coarsening process, referred to as Ostwald ripening, is a
general feature and also occurs between islands in different layers.
The adatom motion between islands of different layers is different
from adatom motion for islands within the same layer due to the
so-called Enrich-Schwoebel (ES) barrier \cite{ehrlich_schwoebel}. Atoms that detach
from the top island have to overcome this additional energy barrier
before they can be incorporated into the bottom island. While the
understanding of Ostwald-ripening processes between islands in the
same layer has advanced significantly in recent years for metals
\cite{morgenstern99_SS,schulze98} as well as for semiconductors
\cite{theis95}, the ripening between islands on different layers has
not been investigated thoroughly, although post-annealing of
three-dimensional films is often used as a simple way of forming
smooth surface layers. Only one specific aspect of the ripening
between layers has attracted a lot of interest \cite{giesen98},
since it has been shown that on Cu(111) and Ag(111) surfaces,
ripening increases by orders of magnitude when the distance between
the island edges falls below a critical width
\cite{giesen98,morgenstern98_prl,giesen99,morgenstern00,giesen00,morgenstern01}.
On Cu(111) it is found that this critical distance coincides with
the distance where the surface state is depopulated. 
For Ag(111), however, the critical distance for the occupation of the surface
state is far above the experimentally determined critical width
\cite{morgenstern00,giesen00,morgenstern01}. Thus, these
observations cannot be explained by quantum size effect and the
underlying mechanism is still under intense debate \cite{larsson01}.

In this Letter we show how lateral electron confinement influences the 
decay of individual adatom islands, resulting in distinct quantum size effects.
We have studied in detail the decay of islands
adsorbed concentrically on top of larger islands. By means of fast-scanning STM, 
we find a transition between attachment-limited decay
for large islands, where the decay is determined by the interface to
the bottom island, to diffusion-limited decay for smaller islands,
where the decay is dominated by the ability of the adatom to diffuse
to the next sink. The shift between these two limiting regimes can
be well understood within an ordinary continuum model and is caused by the
increasing distance between the two step edges of the top and bottom
islands during the decay. 
Very surprisingly, we find that the island decay is reduced
significantly at certain island sizes.
We relate these distinct island sizes to the absence of quantum well states near the Fermi
level and conclude that the energy barrier for adatom detachment depends on the quantum confinement of electrons.
Thus, we present the first observation of the influence of
electron confinement onto surface kinetics.

The experiments were performed on Ag(111) single crystal surfaces,
which were cleaned by 1keV Ar$^+$ sputtering for 30 min and
annealing to $\sim 650^\circ$C for 30 min. Several sputter-anneal
cycles were followed by a final flash to $\sim 850^\circ$C, leading
to terraces with a width larger than 1000 nm. Up to 7 monolayers (ML) of silver was
deposited from an evaporator from Oxford Instruments with a rate
between 0.3 and 0.8 ML/min resulting in stacks of adatom islands
\cite{morgenstern_phd}. The subsequent change in morphology of the
islands during annealing experiments was recorded by means of the
 home-built variable-temperature Aarhus STM and visualized in the form of so-called STM movies, 
i.e.\ sequences of time-lapsed STM images
\cite{Bes96}. As in earlier similar studies special precaution was
taken to avoid any influence of the scanning STM tip
\cite{morgenstern96,morgenstern98_prl}. The microscope is operated
in a standard ultra-high vacuum system (base pressures of $5\cdot
10^{-11}$ Mar) that was equipped with standard surface science
techniques for sample preparation and characterization. The high
thermal stability of the STM combined with a special, active thermal
drift-compensated routine  allowed us to follow dynamically the
morphological changes on surfaces over extended time periods ($>
13$h) \cite{morgenstern96_prl,morgenstern98_prl,morgenstern99_prl}.

\begin{figure}
\includegraphics[width=.45\textwidth]{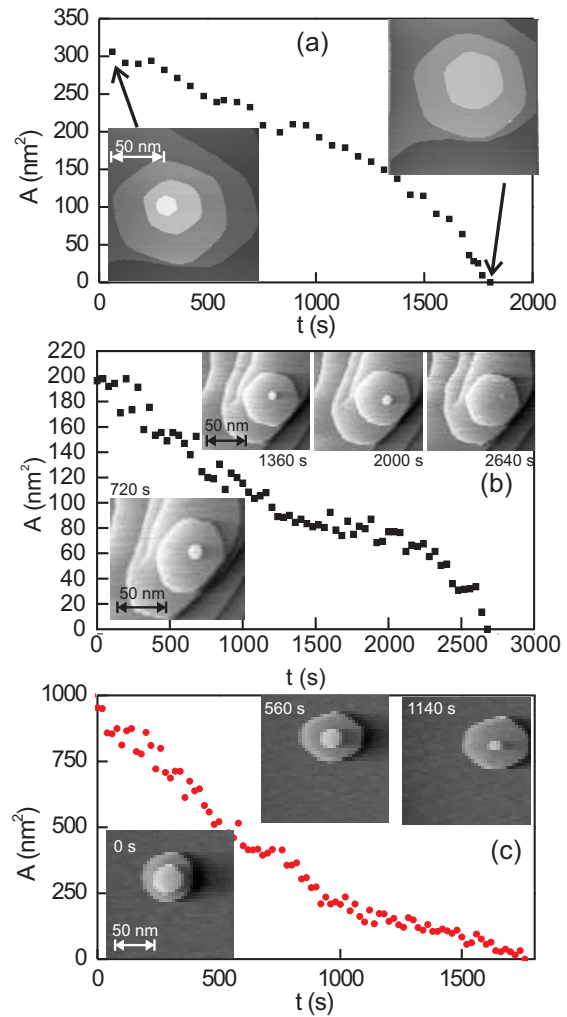}
\caption[]{Decay of top-layer island of an adatom island stack: snapshots
of STM movie and area development (a) 330 K, initial size of bottom
island 2500 nm$^2$, 40 pA, 2 V (b) RT, initial island size of bottom
island: 3500 nm$^2$, 5 nA, 0.3 V (c) RT,  island size of bottom
island changes from 2800 nm$^2$ to 4000 nm$^2$ during top-layer island
decay, 1.5 nA, -1 V. \label{ad_on_ad}}
\end{figure}

After obtaining stable imaging conditions, the dynamic coarsening of
the islands was followed by repeatedly scanning the same spot of the
surface. Figure \ref{ad_on_ad}a shows the decay of the top-layer of
a four-layered Ag island at 330 K. This particular island shows the
decay characteristics expected for island decay in the diffusion
limit: The decay rate smoothly increases with decreasing island
size. 
Often, however, the decay curves exhibit regions in which the
decay proceeds at a slower rate (Fig.\   \ref{ad_on_ad}b and c).
This decrease in decay rates occurs at specific island sizes, e.g.\
for the island in Fig.\ \ref{ad_on_ad}b  at $\approx 80$ nm$^2$. In this
example the lower adatom island gets smaller during the island
decay. Figure \ref{ad_on_ad}c demonstrates that the decreased
decay rates also occur for a lower island, whose area is
increasing.

Before discussing the decreased decay rates, we will discuss the
shape of the decay curves aside from the plateaus. The decay of
adatom islands on terraces
\cite{morgenstern96_prl,morgenstern98_prl} follows a functional
dependence of the island area $A$ on time of $A=a\cdot
(t_0-t)^{2b}$. The decay exponent $b$, obtained by fitting this
power law to the island in Fig.\ \ref{ad_on_ad}a, varies
continuously  from $b=(0.27 \pm 0.01)$ for the total curve over
$(0.29 \pm 0.02)$ from t $>1000$ s to $(0.3\pm 0.1)$ from t $>1500$ s. The
last part of the decay curve for the island in Fig.\ \ref{ad_on_ad}b
has a decay exponent, which increases continuously from 0.2 (considering
the last 1500 s) to $\approx 0.25$ for the last 1000 s to $> 0.3$
for the last 500 s. This  variation in the decay exponent with
island size is in variance with previously observed fixed exponents
for island decay on terraces or within vacancy islands
\cite{morgenstern96_prl,morgenstern99_prl,peale92,wen96,hannon97,ichimiya97,giesen98,morgenstern98_prl}.

\begin{figure}
\includegraphics[width=.35\textwidth]{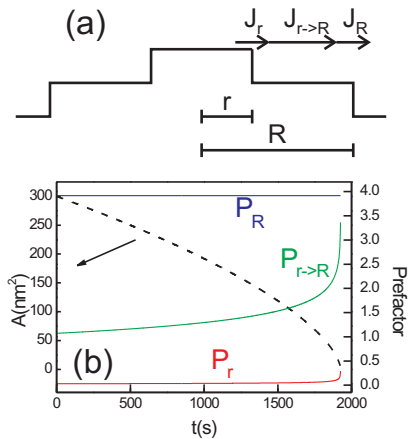}
\caption[]{(a) Theoretically considered scenario, for details see text
(b) Decay curve from
integration for the island in Fig.\ 1a (dashed line) with the value of prefactors as indicated
(full lines);
integration values:
$A_R = 2560$ nm$^2$, $E_s = 0.13$ eV, $E_e=0.71$ eV, $\gamma=0.75/$nm, $T= 330$ K.
\label{theorie}}
\end{figure}

We  account for the varying decay exponent by solving the diffusion equation in an
ordinary continuum description, which has already been used with
success to describe the decay of adatom and vacancy islands within
large vacancy islands and on large terraces
\cite{morgenstern_phd,morgenstern96_prl,morgenstern98_prl}.
In this model, we consider the geometry
sketched in Fig.\ \ref{theorie}a \cite{morgenstern_phd}: A small
adatom island of radius $r$ is placed concentrically on top of a
larger adatom island of radius $R$. The decay of the top-layer island can
be described by the three net fluxes indicated in Fig.\
\ref{theorie}a:  (i) The flux of adatoms from the adatom island to
the terrace $J_r$, (ii) the diffusion flux of adatoms over the
terrace $J_{r\rightarrow R}$, and (iii) the attachment flux of
adatoms to the outer island $J_R$.
Considering mass conservation and equality of net fluxes at steady state
in  analogy to the theoretical modeling of a vacancy island within a vacancy island
\cite{morgenstern98_prl}, yields a differential equation for the
island area $A=\pi r^2$:
\begin{equation}
\label{eq_to_integrate}
\frac{dA}{dt} 
= -\frac{2\pi D}{n}\frac{\rho_{eq}(r)-\rho_{eq}(R)}{P_r+P_{r\rightarrow R}+P_R}
\end{equation}
where $n$ is the atomic density in the surface layer, $D$ is the
diffusivity of adatoms over the surface, 
and
$\rho_{eq}(\hat{r})=\rho_\infty\cdot exp(\frac{\gamma}{kTn\hat{r}})$ 
is the equilibrium adatom concentration of a step of curvature
$1/\hat{r}$ with k the Boltzmann constant and $T$ the absolute temperature. 
The prefactors $P_r= a/r$, $P_{r\rightarrow R}= ln(R/r)$, and $P_R=a/Rs$
result from the three net fluxes involved (Fig.\ \ref{theorie}a) with $a$ the surface lattice constant and 
$s = s_0\cdot e^{-E_s/kT}$, with $E_s$ the ES barrier.

$P_r$ is by far the smallest term for our experimental configuration,
reaching at most a tenth of $P_R$, and it is therefore
negligible ($E_s = 0.13$ eV, $s_0=0.25$ for Ag(111))
\cite{morgenstern98_prl}. The relative importance of the other two
terms changes during the decay. A dominance of $P_R$ is indicative
of an attachment-limited decay, i.e., the decay kinetics is
dominated by the ability of atoms to attach to the bottom island. If,
however,  $P_{r\rightarrow R}$ dominates, the decay is diffusion-limited, 
i.e., the decay kinetics is dominated by the diffusion of
the adatoms between the island edges.

We can linearize the exponential expressions for the adatom
concentrations of eq.\ \ref{eq_to_integrate} and thereby obtain an
analytical solution in the attachment-limited case ($P_R>>
P_{r\rightarrow R}$) which yields $A=a\cdot (t_0-t)^{2b}$ with an invariable
decay exponent of $b=1/3$. However, the increasing importance of the
diffusion-limited term  for decreasing island size results in a
mixture of the two decay regimes, and thus in a variety of
'apparent' decay exponents, in accordance with our experimental
observation. A correct description of the experimental data in
principle demands a numerical integration of eq.\
\ref{eq_to_integrate}, as is shown in Fig.\ \ref{theorie}b
for the experiment of Fig.\ \ref{ad_on_ad}a
using the material parameters of Ag(111) determined
previously \cite{morgenstern98_prl}.
The increasing importance of the diffusion-limited term is evident from the
increasing value of the prefactor $P_{r\rightarrow R}$.

The island size $\tilde{r}$ at which the decay will be predominantly diffusion-limited
($P_R<P_{r\rightarrow R}$)
is given by:
\begin{equation}
\tilde{r}= R\cdot e^{\frac{a}{Rs}}\\
\label{dominance}
\end{equation}
The type of island decay depends thus on the sizes of the two
islands involved ($r$ and $R$) and the Ehrlich-Schwoebel barrier,
i.e.\ the material and specific crystal face.

\begin{figure}
\includegraphics[width=.3\textwidth]{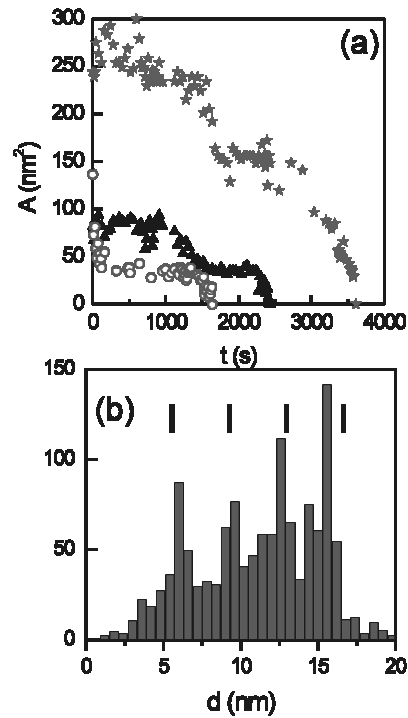}
\caption[]{
(a) Areas of several top-layer islands decaying at RT
(b) Histogram of measured island sizes for 10 different islands; island sizes have been
measured every 10 s; vertical bars indicate island sizes, which have no quantum well state at the Fermi level (see text).
\label{radii}}
\end{figure}

The continuum theory thus explains the observed continuous variation
of the decay exponent. It can not, however, explain the observed
decrease in decay rate at certain distinct island sizes (Fig.\
\ref{ad_on_ad}b). Such apparent decay plateaus are observed for many
islands for lower island sizes varying from 150 to 10.000 nm$^2$ regardless 
of whether the lower island grows or decays. In Fig.\
\ref{radii}a we show some more decay curves. These reveal a distinct
reduction in the decay rate by approximately one order of magnitude for different islands at certain island
sizes. Although the different islands show plateaus at similar values,
not all decay curves show plateaus at all the characteristic island
size values. 

In fig.\ \ref{radii}b we
have plotted the diameter from the decay curves measured at fixed time intervals for several islands
in a histogram.
Distinct preferred sizes are observed for
islands diameters of 6 nm, 9.3 nm, 12.6 nm, and 15.6 nm.
These peaks correspond to those island diameters, where the decay rate is much slower than expected
from the continuum treatment.

For the vertical growth mode reported in \cite{otero02} the island
heights with no quantum well states at the Fermi
level are preferred. In the lateral dimension, magic island sizes without quantum well
states of the surface state electrons near the Fermi level are given
approximately by $\lambda_F/4 + n\cdot \lambda_F /2$. With
$\lambda_F/2 =3.7$ nm for Ag(111), which leads to 5.55, 9.25, 12.95,
and 16.65 nm for n = 1 ... 4. The comparison to the maxima in Fig.\
\ref{radii}b reveals a clear correlation between the size of the
stabilized islands and these values.

Thus, we have shown that the island decay slows down, whenever there is no quantum well state
near the Fermi level. 
This implies that not only the total
energy of an island oscillates as a function of its size, as in ''electronic growth''
\cite{hinch89,otero02,smith96,altfelder97,huang98,gavioli99,su01,zhang98,aballe04},
but also the energy barrier for detachment does so.

From the comparison of the decay rates, in the 'regular' decay regime
and the decreased decay rates we can estimate the energy difference
of this detachment barrier (under the assumption of the same
prefactor) to $(0.07 \pm 0.03)$ eV. This is about one tenth of the
total detachment rate, which was determined in a different decay
experiment to 0.7 eV \cite{morgenstern98_prl}.

In conclusion, the study of the dynamics of the top-layer of a multilayered
adatom island stack on Ag(111) with a fast scanning tunneling microscope reveals two remarkable phenomena:
A transition from attachment-limited to diffusion-limited decay,
and a quantum size effect in island kinetics.
Both phenomena are expected not to be restricted to the particular system.
The former is important whenever the island sizes are close to fulfilling equation \ref{dominance}.
The latter will influence island decay for all surfaces with occupied surface states.

We acknowledge stimulating discussions with Roberto Otero.
One of us (K.M.) acknowledges support by the Deutsche Forschungsgemeinschaft via a Heisenberg scholarship.

\end{document}